\documentstyle[12pt,epsfig]{article}
\begin{document}

\centerline{\Large A definition of the Ponzano-Regge quantum gravity}
\centerline{\Large model in terms of surfaces}
\vspace{0.3cm}
\centerline{\rm Junichi Iwasaki}
\centerline{ Physics Department, University of Pittsburgh,
                  Pittsburgh, PA 15260, USA}
\centerline{\rm E-mail:\ iwasaki@phyast.pitt.edu}
\vspace{0.2cm}
\centerline{April 14, 1995; revised on May 8, 1995}

\begin{abstract}
We show that the partition function of the Ponzano-Regge quantum
gravity model can be written as a sum over surfaces in a $(2+1)$
dimensional space-time.
We suggest a geometrical meaning, in terms of surfaces,
for the (regulated) divergences that appear in the partition function.
\end{abstract}

This paper is dedicated to the 5,100
people killed by the Kobe earthquake
in Japan on January 17, 1995.

\section{Introduction}\label{sec:intro}

\begin{figure}
\centerline{\epsfig{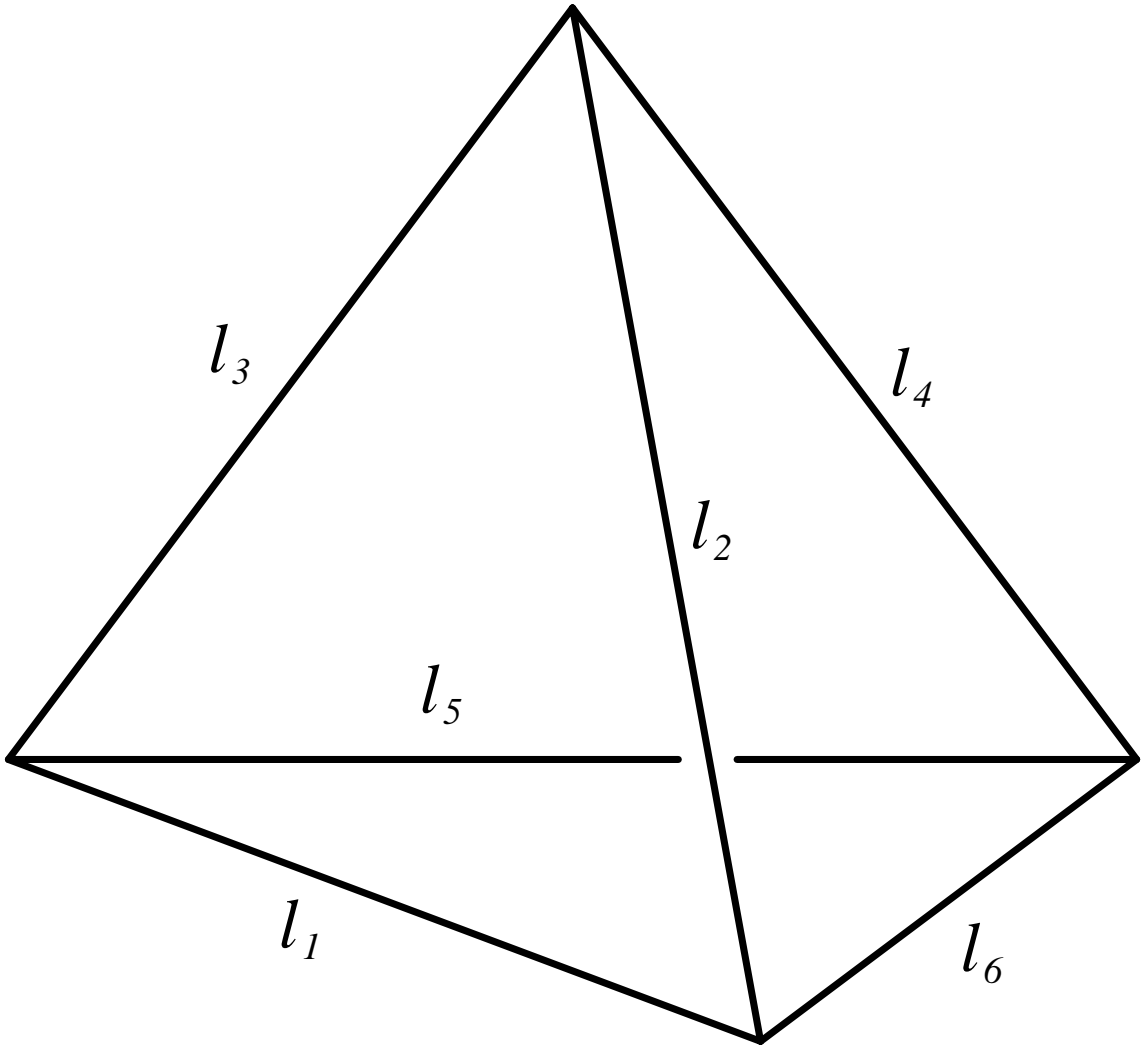}}
\caption{The tetrahedron with edge lengths $l_i\ (i=1,2,\cdots,6)$.}
\label{fig:tetra}
\end{figure}

The Ponzano-Regge quantum gravity model has been a mystery
since the time of its construction \cite{ponzano}.
Ponzano and Regge found an unexpected relation between
the 6j-symbol and Regge's discrete version of
the action functional of
(Euclidean) general relativity  in 3-dimensions.
The description of the relation follows.
Consider a 3-dimensional simplex manifold, which consists of
tetrahedra whose faces are attached to one another.
Pick a tetrahedron and denote the lengths of its edges by
$l_s\ (s=1,2,\cdots,6)$
such that $(l_1,l_2,l_3)$, $(l_3,l_4,l_5)$, $(l_1,l_5,l_6)$
and $(l_2,l_6,l_4)$
form the four faces of the tetrahedron.
(See Fig.~\ref{fig:tetra}.)
The Regge action for the tetrahedron is
\begin{equation}
S_{Regge}=\sum_{s=1}^6 l_s\theta_s,
\end{equation}
where $\theta_s$ is the angle between the outward normals
of two faces sharing the $s$-th edge.
If we assign non-negative half-integer
 values $j_s\ (s=1,2,\cdots,6)$
to the edges such that $l_s=\sqrt{j_s(j_s+1)}$
(or $l_s\approx j_s+{1\over2}$ if $j_s\gg 1$)
and the three $j$'s around every face satisfy triangle relations
(i.e. $|j_1-j_2|\le j_3\le j_1+j_2$;
In this paper, a non-negative half-integer means
an element of the set
$\{0,{1\over2},1,{3\over2},2,\cdots\}$.),
then the 6j-symbol
$\left\{\matrix{j_1&j_2&j_3\cr j_4&j_5&j_6\cr}\right\}$
is non-zero and
 approximates the cosine of the Regge action
in the large $j$ limit.
The positive frequency part of the cosine function is
the integrand of the partition function of quantum (Euclidean)
general relativity
(up to some subtleties concerning the imaginary unit
$i\equiv\sqrt{-1}$) in the pathintegral formalism.
Based on this observation, Ponzano and Regge constructed a
``regularized" partition function of quantum gravity
on the simplex manifold
in terms of 6j-symbols.

By constructing the analogous model
in terms of a quantum group,
Turaev and Viro \cite{turaev} proved that the value of the partition
function of this model is independent of the choice of
the tetrahedral triangulation of 3-manifold.
{}From this fact it could be said that the partition function
of this model is invariant under the renormalization group
transformation and depends  only on the topology of the manifold.
This could be an indication that the model is related to some
continuum quantum theory.

By constructing a Hilbert space of
state functionals on a 2-space
formed and triangulated by the faces of the tetrahedra
in the simplex manifold,
Ooguri \cite{ooguri} showed that this model is isomorphic
to Witten's canonical formulation of $ISO(3)$ Chern-Simons
theory \cite{witten}, which is known to be equivalent to
the Ashtekar formulation of general relativity in $(2+1)$-dimensions
\cite{2+1,abhay}.
(In Ooguri's construction the relevant group is
$SO(3)$ rather than $SO(2,1)$.)

These facts suggest that this model may contain some essential
information of $(2+1)$ quantum gravity.
However, from the model itself,
its geometrical meaning is not transparent
and this ambiguity makes the model confusing
as a physical model.
(Another problem of the model is that the partition function
has ``infrared'' divergences
if the topology of manifold is $\Sigma\times S^1$,
where $\Sigma$ is a closed 2-manifold with genus $g>0$
in addition to the (regulated) divergences addressed in this paper.
\cite{ooguri})
Therefore, it is desired to find a definition of the model
in a geometrically meaningful fashion.

Rovelli \cite{rovelli} pointed out
that Ooguri's representation space is
closely related to the loop representation \cite{loop} of
canonical quantum general relativity. \cite{2+1,abhay}
Recently, it has been suggested that the partition function
of the model may be reconstructed as a sum over surfaces
in such a way that its relation to the loop representation
of canonical $(2+1)$ quantum (Euclidean) general relativity may be
revealed \cite{reform}.

In this paper, we show that
the partition function of the Ponzano-Regge model can be written
as a sum over surfaces and suggest a geometrical meaning,
in terms of surfaces,
for the (regulated) divergences
that appear in the partition function.

In fact, what will be done in this paper is the following.
We present
a partition function which is a sum of products of terms.
Each of the terms is a simple combination of variables associated
with a vertex, an edge, a face or a tetrahedron of
a fixed simplex lattice in a fixed 3-manifold.
The values of these variables are determined by simple
geometrical relations between the simplex lattice
and a family of surfaces.
The computations needed in the evaluation of this partition
function are very simple combinatorial ones
in contrast to the case for the Ponzano-Regge partition function,
which includes also combinatrial but quite complicated
computations of 6j-symbols in addition to their unclear
geometrical meaning.
Nevertheless, this partition function is shown
to be equivalent to the Ponzano-Regge partition function.

Secondly, the regulator introduced by Ponzano and Regge to
regulate divergences appearing in the partition function
is reconstructed by means of the variables defined by
geometrical relations between the simplex lattice and
a family of surfaces and shown to have a geometrical meaning
although the regulator was originally constructed
algebraically by Ponzano and Regge.
We would like to consider this partition function
as a geometrically meaningful definition, in terms of surfaces,
of the Ponzano-Regge model.

In Sec.~\ref{sec:model} we briefly describe
the 6j-symbolic formalism
of the model due to Ponzano and Regge.
In Sec.~\ref{sec:surf} we present the partition function
as a sum over surfaces and show the equivalence
to the 6j-symbolic form of the partition function.
In Sec.~\ref{sec:div}
we suggest a geometrical meaning, in terms of surfaces,
for the (regulated)
divergences that appear in the partition function.
We conclude our work in Sec.~\ref{sec:concl}.

\section{The Ponzano-Regge model}\label{sec:model}

The Ponzano-Regge model is defined by the partition function
described as follows.
Fix a closed 3-manifold $M$.
Introduce a tetradedral triangulation $T$
in $M$.
Let $S_0$, $S_1$, $S_2$ and $S_3$ denote the sets of the vertices,
edges, faces and tetrahedra in $T$ respectively.
Let $N_0$, $N_1$, $N_2$ and $N_3$ denote the numbers of elements
of $S_0$, $S_1$, $S_2$ and $S_3$ respectively.
Assign a non-negative half-integer $j_e$ to each $e\in S_1$.
Assign a 6j-symbol to each tetrahedron $t\in S_3$
with six non-negative half-integers
$j_s (s=1,2,\cdots 6)$ already assigned
to the edges of $t$ in the way that
$(j_1,j_2,j_3)$, $(j_3,j_4,j_5)$, $(j_1,j_5,j_6)$ and $(j_2,j_6,j_4)$
are the edges of respective faces of $t$.
Then the partition function is defined by \cite{ponzano,turaev}
\begin{equation}
Z_{PR}(M,T):=\lim_{N\to\infty}\sum_{j\le N}
\Lambda^{-N_0}(N)\prod_{e\in S_1}(-1)^{2j_e}(2j_e+1)
\prod_{t\in S_3}
(-1)^{-\sum_sj_s}\left\{\matrix{j_1 & j_2 & j_3\cr
j_4 & j_5 & j_6} \right\},
\label{eq:part}
\end{equation}
where sums are taken for all $j$'s of edges
over all the non-negative half-integers less than or equal to $N$.
Here $\Lambda(N)$ is introduced to regulate divergences
and is defined by
\begin{equation}
\Lambda(N):=\sum_{k=0,{1\over2},1,\cdots N}(2k+1)^2,
\label{eq:lambda}
\end{equation}
which behaves as
$\Lambda(N)\sim{4\over3}N^3$ as $N\to\infty$.
If the $j$'s of the three edges of any face do not satisfy
the triangle relations (i.e. $|j_1-j_2|\le j_3\le j_1+j_2$),
then we define the 6j-symbol including these edges to vanish and hence
the partition function has many vanishing terms.
The variables of this model are the lengths of edges
$l=j+{1\over2}$
or $j$ restricted to non-negative half-integers.
The way of gluing the tetrahedra, in other words
the connectivity and the number of tetrahedra,
is fixed in this model.

\section{The partition function in terms of surfaces}
\label{sec:surf}

We claim that the partition function of the model for the fixed
3-manifold $M$ with the tetrahedral triangulation $T$ can be
written as
\begin{eqnarray}
&Z_S(M,T)=&\lim_{N\to\infty}\sum_{j\le N}\sum_{K\le N}
\sum_{A,B\le N}
\left(\prod_{v\in S_0}\Lambda^{-1}(N)\right)
\left(\prod_{e\in S_1}{\cal L}_e(j_e)\right)
\times
\nonumber\\&&
\left(\prod_{f\in S_2}{\cal A}_f^{-1}(K^f)
\prod_{\scriptstyle i,i'\atop\scriptstyle (i<i')}
\delta[K_i^f+K_{i'}^f,j_{e(f,i,i')}]\right)
\times
\nonumber\\&&
\left(\prod_{t\in S_3}{\cal V}_t(A^t,B^t)
\prod_{m,n}\delta[A_m^t+B_n^t,K_{i(t,m,n)}^{f(t,m,n)}]\right),
\label{eq:surf}
\end{eqnarray}
with
\begin{eqnarray}
&&{\cal L}_e:=(-1)^{2j_{e}}(2j_{e}+1),\label{eq:L}\\
&&{\cal A}_f:={(\sum_i2K_i^f+1)!\over\prod_i(2K_i^f)!},\label{eq:A}\\
&&{\cal V}_t:=(-1)^{-\sum_mA_m^t+\sum_n2B_n^t}{\ }
{(\sum_m2A_m^t+\sum_n2B_n^t+1)!\over
\prod_m(2A_m^t)!\prod_n(2B_n^t)!}.\label{eq:V}
\end{eqnarray}
The notation in the equations (\ref{eq:surf}) through
(\ref{eq:V}) is as follows.
(Also see Fig.~\ref{fig:note}.)
The $i$ (or $i'$) is associated with a doublet of edges
of a face $f\in S_2$
and takes the values $1$, $2$ and $3$.
The $m$ is associated with a triplet of edges sharing a single vertex
of a tetrahedron $t\in S_3$ and takes the values $1$, $2$, $3$ and $4$.
The $n$ is associated with a quartet of edges closing a rectangle of
a tetrahedron $t\in S_3$ and takes the values $1$, $2$ and $3$.
The $e(f,i,i')$ means the edge common to the two doublets $i$ and $i'$
of edges of the face $f\in S_2$.
The $i(t,m,n)$ means the doublet of edges common to the triplet $m$
and the quartet $n$ of edges of the tetrahedron $t\in S_s$.
The $f(t,m,n)$ means the face on which $i(t,m,n)$ is a doublet of edges.
The $j_e$, $K_i^f$, $A_m^t$ and $B_n^t$ are non-negative half-integer
variables, assigned respectively to
an edge $e\in S_1$, a doublet $i$ of edges of a face $f\in S_2$,
a triplet $m$ of edges of a tetrahedron $t\in S_3$ and
a quartet $n$ of edges of a tetrahedron $t\in S_3$
and their sums are taken independently for all $e$, $f$, $i$,
$t$, $m$ and $n$ over all the non-negative half-integers
less than or equal to $N$.
The $\delta$'s are the Kronecker delta symbols, that is,
\begin{equation}
\delta[x,y]=\left\{\matrix{1{\rm\ \ \ if\ \ \ }x=y\cr
                           0{\rm\ \ \ if\ \ \ }x\ne y}\right..
\end{equation}

The geometrical meanings of $A_m^t$ and $B_n^t$ are as follows.
Pick a tetrahedron $t\in S_3$.
Imagine a surface inside the tetrahedron on which no vertex lies
and which divides the tetrahedron into two connected regions.
The surface may be closed or open with a boundary
on the faces.
The surface may intersect some of the edges if it is open.
If the surface intersects a triplet of edges sharing a single
vertex, then we say the surface is in the class $m$ of type-$A$,
where $m$ takes the values $1$, $2$, $3$ and $4$ corresponding to
the triplets of edges defined above.
(See Fig.~\ref{fig:surf}.)
If the surface intersects a quartet of edges closing a rectangle,
then we say the surface is in the class $n$ of type-$B$,
where $n$ takes the values $1$, $2$, and $3$ corresponding to
the quartets of edges defined above.
(See Fig.~\ref{fig:surf}.)
Then, given non-negative half-integers  $A_m^t$ and $B_n^t$,
we interpreted them that the tetrahedron $t$ contains
$2A_m^t$ open surfaces of the class $m$ of type-$A$ and
$2B_n^t$ open surfaces of the class $n$ of type-$B$.
We call the surfaces in these classes the {\it cross-sections}
of the tetrahedron $t$.

The geometrical meaning of $K_i^f$ is as follows.
Pick a face $f\in S_2$.
Imagine a curve on the face on which no vertex lies and
which divides the face into two connected areas.
The curve may be closed or open with two ends
at the edges of the face.
If the curve has two ends at two different edges,
then we say the curve is in the class $i$,
where $i$ takes the values $1$, $2$ and $3$ corresponding to
the doublets of edges defined above.
(See Fig.~\ref{fig:surf}.)
Then, given a non-negative half-integer $K_i^f$,
we interpret it that the face $f$ contains
$2K_i^f$ open curves of the class $i$.

The geometrical meaning of $j_e$ is as follows.
Pick an edge $e\in S_1$.
Imagine a point on the edge which does not coincide with any vertex.
(See Fig.~\ref{fig:surf}.)
Then, given a non-negative half-integer $j_e$,
we interpret it that the edge $e$ contains $2j_e$ points.
(Two points may coincide in position.)

Now,  we show that the partition function Eq.~(\ref{eq:surf})
can be understood as a sum over (closed) surfaces.
First, given a set of non-negative half-integers
$\{j_e,K_i^f,A_m^t,B_n^t\}$ for all
$e\in S_1$, $i$ on $f\in S_2$, and $m$ and $n$ in $t\in S_3$,
place $2j_e$ points on $e\in S_1$, $2K_i^f$ curves of the class $i$
on $f\in S_2$ and $2A_m^t$ cross-sections of the class $m$
of type-$A$ and $2B_n^t$ cross-sections of the class $n$
of type-$B$ in $t\in S_3$.
By moving the points on edges and/or deforming the curves on faces
and the cross-sections in tetrahedra
keeping their classes unchanged,
we can connect all of them
(such that cross-sections are connected by curves and
curves are connected by points
like in Fig.~\ref{fig:surf})
to form a family of closed surfaces
in the manifold $M$ if the set of half-integers satisfies
the conditions given by the non-zero Kronecker delta symbols
in Eq.~(\ref{eq:surf}).
Note that there is no vertex on any of the family of closed surfaces.

Second, suppose a family of closed surfaces on which no vertex
lies is given in the manifold $M$.
Some of the surfaces may intersect an edge $e\in S_1$.
Assign to $j_e$ half of the number of times the surfaces
intersect the edge $e$.
Some of the surfaces may intersect a face $f\in S_2$ and
form curves of the class $i$.
Assign to $K_i^f$ half of the number of curves of the class $i$
formed on the face $f$.
Some of the surfaces may intersect a tetrahedron $t\in S_3$
and form cross-sections.
Assign to $A_m^t$ half of the number of cross-sections
of the class $m$ of type-$A$ formed in the tetrahedron $t$.
Assign to $B_n^t$ half of the number of cross-sections
of the class $n$ of type-$B$ formed in the tetrahedron $t$.
Therefore, given a family of closed surfaces on which
no vertex lies,
a set of non-negative half-intergers $\{j_e,K_i^f,A_m^t,B_n^t\}$
is determined.

{}From the two facts above, we can replace the sums
over non-negative half-integers $\{j_e,K_i^f,A_m^t,B_n^t\}$
in Eq.~(\ref{eq:surf}) by
a sum over families of closed surfaces on which
no vertex lies such that every non-negative half-interger in
$\{j_e,K_i^f,A_m^t,B_n^t\}$ determined by a family
is less than or equal to $N$.

However, some families do not satisfy the conditions given by
the non-zero Kronecker delta symbols and hence give rise to
vanishing terms in the partition function.
We call the closed surfaces of which the families determining
non-vanishing terms consist {\it spin-surfaces}.
This name is based on our interpretation that
angular momentum value $j_e$ of an edge $e\in S_1$
in the 6j-symbolic formulation of the model is determined
by the fact that the edge is intersected (or ``interacted")
$2j_e$ times by the surfaces as if each of them has spin-$1\over2$.

Moreover, some families of spin-surfaces may produce
an identical set of non-negative half-integers
$\{j_e,K_i^f,A_m^t,B_n^t\}$.
We define equivalence classes of families of spin-surfaces and say
two families of spin-surfaces are equivalent
if they produce an identical set
of non-negative half-integers $\{j_e,K_i^f,A_m^t,B_n^t\}$.
For example, a family of spin-surfaces can be deformed
or cut and reconnected in a different way
such that its equivalence class is unchanged.
(In this sense, intersections of the surfaces do not play any role.)
Therefore,  the partition function can be expressed as a sum over
equivalence classes of families of spin-surfaces
(or equivalence classes of spin-surface families in short)
such that every non-negative half-integer
in $\{j_e,K_i^f,A_m^t,B_n^t\}$
determined is less than or equal to $N$.
We can write the partition function formally as
\begin{equation}
Z_S(M,T)=\lim_{N\to\infty}
\sum_{\scriptstyle\{\sigma\}_N}
\prod_{v\in S_0}\Lambda^{-1}(N)
\prod_{e\in S_1}{\cal L}_e
\prod_{f\in S_2}{\cal A}_f^{-1}
\prod_{t\in S_3}{\cal V}_t,
\label{eq:spsf}
\end{equation}
where $\{\sigma\}_N$ stands for the equivalence classes of spin-surface
families with a condition that every non-negative half-integers
determined from them,
$\{j_e(\sigma),K_i^f(\sigma),A_m^t(\sigma),B_n^t(\sigma)\}$,
is less than or equal to $N$.
Note that this partition function consists of only combinatorial
computations.

The next task is to show that our partition function Eq.~(\ref{eq:surf})
is in fact equivalent to the 6j-symbolic form of the partition function
Eq.~(\ref{eq:part}).
First, insert
\begin{equation}
\prod_{t\in S_3}\prod_{s=1}^{6}\sum_{J_s^t\le N}
\delta[J_s^t,\sum_{m(s)}A_m^t+\sum_{n(s)}B_n^t]=1
\label{eq:delta}
\end{equation}
into Eq.~(\ref{eq:surf}), where the sums are taken for all $s$ and $t$
independently over all the non-negative half-integers
less than or equal to $N$
and $A_m^t\le N$ and $B_n^t\le N$.
The $s$ is associated with six edges of the tetrahedron $t\in S_3$.
The $m(s)$ means the two triplets $m$'s of edges
(of the tetrahedron $t$)
having the edge $s$ and the $n(s)$ means the two quartets $n$'s
of edges (of the tetrahedron $t$) having the edge $s$.
{}From the conditions given by the non-zero Kronecker delta symbols
in Eq.~(\ref{eq:delta}), we can write $A_m^t+B_n^t$ in terms of
$J_s^t$'s, that is,
\begin{equation}
2(A_m^t+B_n^t)=\left(\sum_{s(t,m,n)}J_s^t\right)-J_{\bar s(t,m,n)}^t,
\end{equation}
where the $s(t,m,n)$ means the two edges in the doublet $i(t,m,n)$
and the $\bar s(t,m,n)$ means the edge of $f(t,m,n)$
not in the $i(t,m,n)$.
The $s(t,m,n)$'s and $\bar s(t,m,n)$ form a triangle (a face).

Then, Eq.~(\ref{eq:surf}) becomes
\begin{eqnarray}
&&Z_S(M,T)=\lim_{N\to\infty}\sum_{j\le N}\sum_{K\le N}
\sum_{A,B\le N}\sum_{J\le N}
\left(\prod_{v\in S_0}\Lambda^{-1}(N)\right)
\left(\prod_{e\in S_1}{\cal L}_e(j_e)\right)
\times
\nonumber\\&&
\left(\prod_{f\in S_2}{\cal A}_f^{-1}(K^f)
\prod_{\scriptstyle i,i'\atop\scriptstyle (i<i')}
\delta[K_i^f+K_{i'}^f,j_{e(f,i,i')}]\right)
\times
\nonumber\\&&
\left(\prod_{t\in S_3}{\cal V}_t(A^t,B^t)
\prod_{m,n}\delta[\sum_{s(t,m,n)}J_s^t-J_{\bar s(t,m,n)}^t,
2K_{i(t,m,n)}^{f(t,m,n)}]
\prod_{s=1}^6\delta[J_s^t,\sum_{m'(s)}A_{m'}^t
+\sum_{n'(s)}B_{n'}^t]\right).
\label{eq:much}
\end{eqnarray}
Notice that the first Kronecker delta symbol in the last parentheses
can be rewritten as
\begin{eqnarray}
&&\prod_{t\in S_3}
\prod_{m,n}\delta[\sum_{s(t,m,n)}J_s^t-J_{\bar s(t,m,n)}^t,
2K_{i(t,m,n)}^{f(t,m,n)}]
=\prod_{f\in S_2}
\prod_{t(f)}\prod_i\delta[\sum_{s(f,i)}J_s^t-J_{\bar s(f,i)}^t,
2K_{i}^{f}]
\nonumber\\&&
=\prod_{f\in S_2}
\prod_{t(f)}\prod_{\scriptstyle i,i'\atop\scriptstyle (i<i')}
\delta[J_{s(f,i,i')}^t,K_{i}^{f}+K_{i'}^{f}],
\end{eqnarray}
where the $t(f)$ denotes the two adjacent tetrahedra of the face $f$,
the $s(f,i)$ means the two edges (of the tetrahedron $t$)
in the doublet $i$ of the face $f$,
the $\bar s(f,i)$ means the edge (of the tetrahedron $t$)
of the face $f$ not in the doublet $i$
of the face $f$, and
the $s(f,i,i')$ means the edge (of the tetrahedron $t$)
common to the two doublets $i$ and $i'$ of edges of the face $f$.

By suming over $K_i^f$ for all $i$ and $f$,
Eq.~(\ref{eq:much}) becomes
\begin{eqnarray}
&&Z_S(M,T)=\lim_{N\to\infty}\sum_{j\le N}
\sum_{J\le N}
\left(\prod_{v\in S_0}\Lambda^{-1}(N)\right)
\left(\prod_{e\in S_1}{\cal L}_e(j_e)\right)
\times
\nonumber\\&&
\sum_{A,B\le N}
\prod_{t\in S_3}
\left(\prod_{f(t)}{\cal A}_f^{-1/2}(K^f(J^t))\right)
{\cal V}_t(A^t,B^t)
\prod_{s=1}^6\delta[J_s^t,\sum_{m(s)}A_{m}^t
+\sum_{n(s)}B_{n}^t]{\ }
\delta[J_s^t,j_{e(t,s)}],
\label{eq:more}
\end{eqnarray}
where the $f(t)$ denotes the four adjacent faces
of the tetrahedron $t$
and the $e(t,s)$ is the edge corresponding to the edge $s$
of the tetrahedron $t$ and
$K^f(J^t)$ is such that
\begin{equation}
2K_{i(t,m,n)}^f=\left(\sum_{s(t,m,n)}J_s^t\right)
-J_{\bar s(t,m,n)}^t.
\end{equation}

The sum of the $A_m^t$ and $B_n^t$ dependent part
for each $t\in S_3$ in Eq.~(\ref{eq:more}) is
\begin{eqnarray}
&&\sum_{A^t,B^t\le N}
\left(\prod_{f(t)}{\cal A}_f^{-1/2}(K^f(J^t))\right)
{\cal V}_t(A^t,B^t)
\prod_{s=1}^6\delta[J_s^t,\sum_{m(s)}A_{m}^t
+\sum_{n(s)}B_{n}^t]
\nonumber\\
&&=\sum_{A^t,B^t\le N}
(-1)^{-\sum_mA_m^t+\sum_n2B_n^t}
\left[{\prod_m\prod_n(2A_m^t+2B_n^t)!\over
\prod_m(\sum_{m'\ne m}2A_{m'}^t+\sum_n2B_n^t+1)!}\right]^{1/2}
\times
\nonumber\\
&&{(\sum_m2A_m^t+\sum_n2B_n^t+1)!\over
\prod_m(2A_m^t)!\prod_n(2B_n^t)!}
\prod_{s=1}^6\delta[J_s^t,\sum_{m(s)}A_{m}^t
+\sum_{n(s)}B_{n}^t]
\nonumber\\
&&=(-1)^{-\sum_sJ_s^t}
\left\{\matrix{J_1^t&J_2^t&J_3^t\cr J_4^t&J_5^t&J_6^t\cr}\right\}.
\label{eq:surprise}
\end{eqnarray}
This is a non-zero 6j-symbol.
The last step in Eq.~(\ref{eq:surprise}) may be a surprise,
but it agrees with the explicit
expression of the 6j-symbol \cite{6-j}, which is
\begin{eqnarray}
&\left\{\matrix{j_1&j_2&j_3\cr j_4&j_5&j_6\cr}\right\}=&
(-1)^{j_1+j_2+j_4+j_5}\Delta(j_1,j_2,j_3)\Delta(j_1,j_5,j_6)
\Delta(j_2,j_4,j_6)\Delta(j_5,j_4,j_3)
\times
\nonumber\\
&&\sum_z(-1)^z (j_1+j_2+j_4+j_5+1-z)!
\left[z!(j_3+j_6-j_1-j_4+z)!\right.
\times
\nonumber\\
&&(j_3+j_6-j_2-j_5+z)!(j_1+j_2-j_3-z)!(j_4+j_5-j_3-z)!
\times
\nonumber\\
&&\left.(j_1+j_5-j_6-z)!(j_2+j_4-j_6-z)!\right]^{-1},
\end{eqnarray}
where
\begin{equation}
\Delta(a,b,c):=\left[{(a+b-c)!(a+c-b)!(b+c-a)!\over
(a+b+c+1)!}\right]^{1/2}
\end{equation}
with $a$, $b$ and $c$ satisfying the triangle relations,
otherwise $\Delta(a,b,c)=0$,
and $z$ runs over integer values which do not lead
to negative factorials.

Finally, by substituting Eq.~(\ref{eq:surprise}) into
Eq.~(\ref{eq:more}), we find
\begin{eqnarray}
&Z_S(M,T)=&\lim_{N\to\infty}\sum_{j\le N}
\sum_{J\le N}
\left(\prod_{v\in S_0}\Lambda^{-1}(N)\right)
\left(\prod_{e\in S_1}{\cal L}_e(j_e)\right)
\times
\nonumber\\&&
\left(\prod_{t\in S_3}
(-1)^{-\sum_sJ_s^t}
\left\{\matrix{J_1^t&J_2^t&J_3^t\cr J_4^t&J_5^t&J_6^t\cr}\right\}
\prod_{s=1}^6\delta[J_s^t,j_{e(t,s)}]\right).
\label{eq:most}
\end{eqnarray}
This is another expression of Eq.~(\ref{eq:part}).

Therefore, our partition function in terms of surfaces
is equivalent to the 6j-symbolic partition function of the model.
Advantages of the use of the surface formalism may be that
it may provide a geometrical interpretation
of the model in terms of surfaces
and that the (regulated) divergences that appear
in the partition function may be
geometrically understood in terms of surfaces.
The first point is deserved for further investigations and,
in particular, it is important to clarify what kind of surfaces
the spin-surfaces are.
The second point is discussed in the next section.

\section{Divergences}\label{sec:div}

In the 6j-symbolic partition function Eq.~(\ref{eq:part}),
$\Lambda(N)$ is present to regulate divergences.
In the surface formalism described in the previous section,
the same regulator $\Lambda(N)$ is used although its meaning
has not yet been explained.
In this section we analyze the regulator in terms surfaces
and suggest a geometrical meaning for the divergences regulated.

\begin{figure}
\centerline{\epsfig{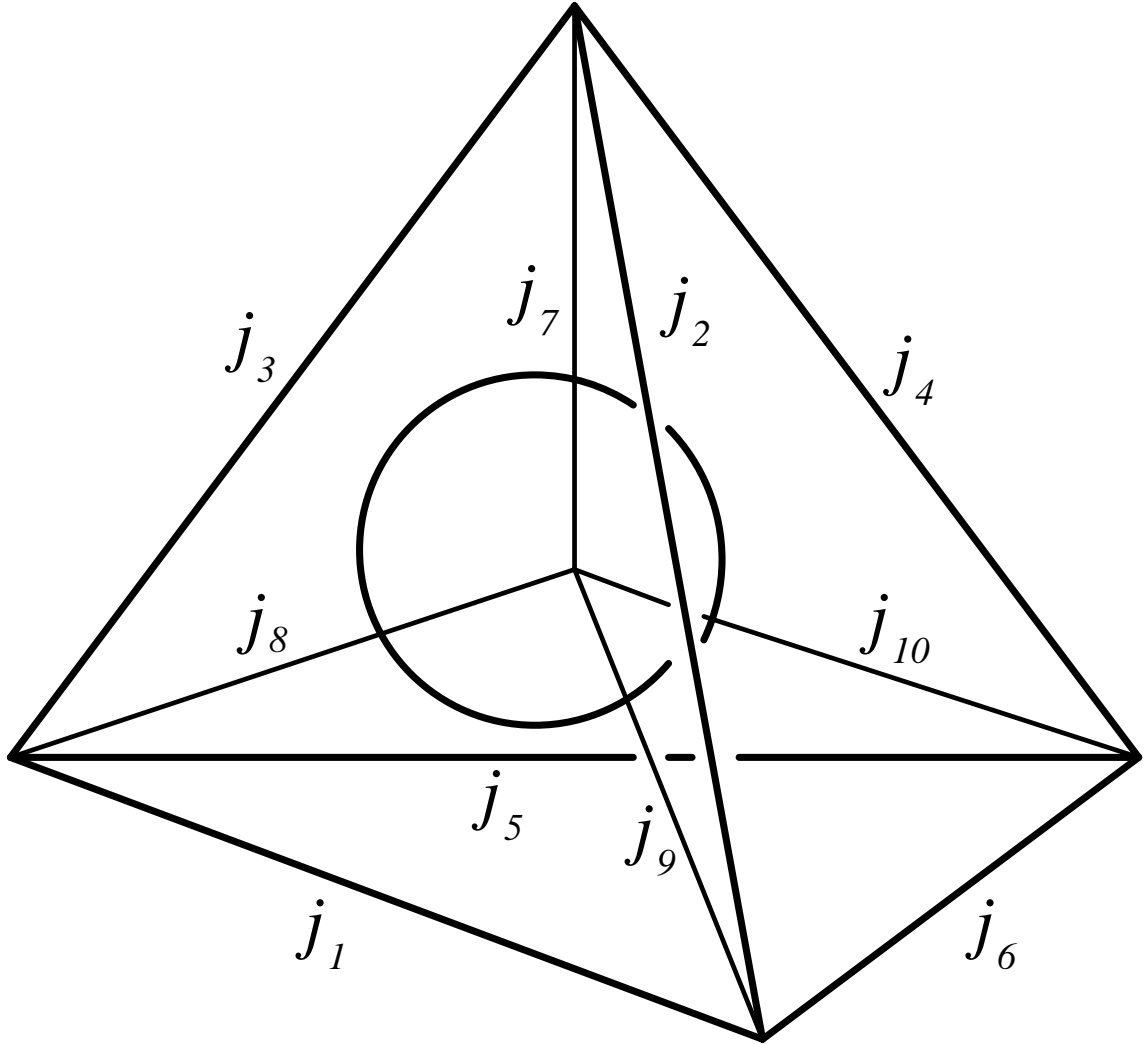}}
\caption{The decomposition of a tetrahedron
to four smaller tetrahedra, and a spin-surface
contributing to $\Lambda(N)$.}
\label{fig:refine}
\end{figure}

Ponzano and Regge introduced the regulator $\Lambda(N)$
defined by Eq.~(\ref{eq:lambda})
in the following way. \cite{ponzano}
Pick a tetrahedron from the tetrahedral triangulation $T$.
By placing one more vertex and four edges in this tetrahedron,
make a finer triangulation consisting of four tetrahedra
in the original tetrahedron and the other tetrahedra in $T$.
(See Fig.~\ref{fig:refine}.)
Assign a non-negative half-integer $j_s$ or $j_a$
to each of the ten edges
of the four tetrahedra.
The $s=1,2,\cdots, 6$ denote the edges on the boundary,
consisting of the four faces of the original tetrahedron, and
the $a=7,8,9,10$ denote the interior edges added.
In addition to the four faces of the original tetrahedron,
there are six more faces formed by
$(j_1,j_9,j_8)$, $(j_2,j_7,j_9)$, $(j_3,j_8,j_7)$,
$(j_4,j_7,j_{10})$, $(j_5,j_8,j_{10})$,
and $(j_6,j_{10},j_9)$.
Assign a 6j-symbol to each of the four tetrahedra.
Then the 6j-symbols satisfy the relation
\begin{eqnarray}
&&\sum_{j_7,j_8,j_9,j_{10}}
\left(\prod_{a=7}^{10}(-1)^{2j_{a}}(2j_{a}+1)\right)
(-1)^{-\sum_{s=1}^{6}2j_s-\sum_{a=7}^{10}3j_{a}}
\times
\nonumber\\
&&\left\{\matrix{j_1&j_2&j_3\cr j_7&j_8&j_9\cr}\right\}
\left\{\matrix{j_6&j_4&j_2\cr j_7&j_9&j_{10}\cr}\right\}
\left\{\matrix{j_5&j_3&j_4\cr j_7&j_{10}&j_8\cr}\right\}
\left\{\matrix{j_1&j_5&j_6\cr j_{10}&j_9&j_8\cr}\right\}
\nonumber\\
&=&(-1)^{-\sum_{s=1}^{6}j_s}
\left\{\matrix{j_1&j_2&j_3\cr j_4&j_5&j_6\cr}\right\}
\left(\lim_{N\to\infty}\Lambda(N)\right),
\label{eq:refine}
\end{eqnarray}
where the sums are taken over all the non-negative half-integers.
Because of this relation, the partition function is made invariant
under the change of tetrahedral triangulations.
The $\Lambda(N)$ is divergent as $N\to\infty$ and the presence of
$\Lambda^{-1}(N)$ for the vertex added in the partition function
divides the divergence away.

However, from Eq.~(\ref{eq:refine}) of the 6j-symbolic formalism,
the geometrical meaning of $\Lambda(N)$ is unclear
since no variable like
$k$ in Eq.~(\ref{eq:lambda}) is assigned to any vertex in the model.
In order to find a geometrical meaning of $\Lambda(N)$
in the surface formalism, consider the case that
$j_s (s=1,2,\cdots 6)$ on the boundary vanish
(note $\left\{\matrix{0&0&0\cr 0&0&0\cr}\right\}=1$)
and  rewrite the 6j-symbols in terms of surfaces
using Eq.~(\ref{eq:surprise}).
Then the Eq.~(\ref{eq:refine}) becomes
\begin{eqnarray}
&&\lim_{N\to\infty}\Lambda(N)
\nonumber\\
&&=\sum_{j_7,j_8,j_9,j_{10}}
\left(\prod_{a=7}^{10}(-1)^{2j_{a}}(2j_{a}+1)\right)
(-1)^{-\sum_{a=7}^{10}3j_{a}}
\times
\nonumber\\
&&\left\{\matrix{0&0&0\cr j_7&j_8&j_9\cr}\right\}
\left\{\matrix{0&0&0\cr j_7&j_9&j_{10}\cr}\right\}
\left\{\matrix{0&0&0\cr j_7&j_{10}&j_8\cr}\right\}
\left\{\matrix{0&0&0\cr j_{10}&j_9&j_8\cr}\right\}
\nonumber\\
&&=\lim_{N\to\infty}\sum_{k\le N}
\left(\prod_e(-1)^{2k}(2k+1)\right)
\left(\prod_f(2k+1)^{-1}\right)
\left(\prod_t(-1)^{-k}(2k+1)\right),
\label{eq:interior}
\end{eqnarray}
where $k$ is a non-negative half-integer variable determined by
equivalence classes of spin-surface families
with spin-surfaces enclosing the interior vertex, intersecting
the four interior edges and forming a cross-section of type-$A$
in each of the four interior tetrahedra
(See Fig.~\ref{fig:refine};
there are only equivalence classes of spin-surface families
consisting of these spin-surfaces
when $j_s$'s on the boundary are zero.)
and $e$, $f$ and $t$ stand for the interior edges,
the interior faces and the interior tetrahedra respectively.

By letting $n_0$, $n_1$ and $n_2$ denote the numbers of
the interior edges,
the interior faces and the interior tetrahedra respectively
and using the relation $2n_1=3n_2$,
Eq.~(\ref{eq:interior}) can be rewritten as
\begin{eqnarray}
&\Lambda(N)&=\sum_{k\le N}
\left[(-1)^{2k}(2k+1)\right]^{n_0}
(2k+1)^{-n_1}
\left[(-1)^{-k}(2k+1)\right]^{n_2}
\nonumber\\
&&=\sum_{k\le N}(-1)^{2k(n_0-n_1+n_2)}{\ }(2k+1)^{n_0-n_1+n_2}
\nonumber\\
&&=\sum_{k\le N}\left[(-1)^{2k}{\ }(2k+1)\right]^{\chi},
\label{eq:lchi}
\end{eqnarray}
where $\chi:=n_0-n_1+n_2=2(1-g)$ is the Euler characteristic
of a closed surface with the genus $g$
embedded in a 3-space.
In our case here $n_0=4$, $n_1=6$, $n_2=4$ and hence $g=0$ and
$\chi=2$.
Therefore, the $\Lambda(N)$ defined in Eq.~(\ref{eq:lambda})
has been recovered here with geometrical meanings:
(i) The $k$ is
the variable determined by equivalence classes of families
of spin-surfaces enclosing only one vertex with the topology
of a 2-sphere (See Fig.~\ref{fig:refine};
$2k$ is the number of such spin-surfaces), and
(ii) the exponent in the right hand side of
Eq.~(\ref{eq:lambda}) is the Euler characteristic of
each of these spin-surfaces.

Note that Eq.(\ref{eq:lchi}) is divergent as $N\to\infty$
if $\chi=2$ or $0$.
In other words, surfaces with $\chi=2$ and $0$ may be
origins of divergences.
This conjecture may give a hint to clarifying the spin-surfaces.

\section{Conclusions}\label{sec:concl}

In this paper, we showed that the partition function of
the Ponzano-Regge quantum gravity model can be written
as a sum over surfaces
in a $(2+1)$ dimensional space-time.
We called the surfaces playing roles in the partition function
{\it spin-surfaces} because they were interpreted to determine
the angular momentum values of edges in the 6j-symbolic form of
the partition function by intersecting (or ``interacting")
the edges as if each surface has spin-$1\over2$.
We analyzed the regulator introduced
by Ponzano and Regge by means of the surface formalism
and suggested a geometrical meaning, in terms of surfaces,
for the (regulated) divergences
that appear in the partition function.

We may think that the existences of this surface formulation
of the model and this way of understanding
the (regulated) divergences that appear
 in the partition function may provide a further
interest to the model.
It may be hoped that this surface formulation may provide
a geometrical way of investigating the otherwise very
subtle model and hence another way of studying well explored problems
in $(2+1)$ quantum gravity from a slightly different angle.

\section*{Acknowledgements}

The author has benefited from discussions with Carlo Rovelli.
The author also thanks Lee Smolin for helpful suggestions.

\filbreak

\filbreak
\begin{figure}
\centerline{\epsfig{figure=fig2.ps,width=150mm}}
\caption{Nomenclature for the partition function.}
\label{fig:note}
\end{figure}

\filbreak
\begin{figure}
\centerline{\epsfig{figure=fig3.ps,width=150mm}}
\caption{Points on edges, curves on faces and
cross-sections in tetrahedra.}
\label{fig:surf}
\end{figure}

\filbreak

\end{document}